# Validation and Administration of a Conceptual Survey on the Formalism and Postulates of Quantum Mechanics

Emily Marshman[1] and Chandralekha Singh[2]

[1]*Department of Physics, Community College of Allegheny County, Pittsburgh, PA*
[2]*Department of Physics and Astronomy, University of Pittsburgh, Pittsburgh, PA*

**Abstract.** We developed and validated a conceptual survey that focuses on the formalism and postulates of quantum mechanics covered in upper-level undergraduate quantum mechanics courses. The concepts included in the Quantum Mechanics Formalism and Postulate Survey (QMFPS) focus on Dirac notation, the Hilbert space, state vectors, physical observables and their corresponding Hermitian operators, compatible and incompatible observables, quantum measurement, time-dependence of quantum states and expectation values, and spin angular momenta. Here we describe the validation and administration of the survey, which has been administered to over 400 upper-level undergraduate and graduate students from six institutions. The QMFPS is valid and reliable for use as a low-stakes test to measure the effectiveness of instruction in an undergraduate quantum mechanics course that covers relevant content. The survey can also be used by instructors to identify students' understanding of the formalism and postulates of quantum mechanics at the beginning and end of a graduate quantum mechanics course since graduate students are expected to have taken an undergraduate quantum mechanics course that covers the content included in the survey. We found that undergraduate students who engaged with research-validated learning tools performed better than students who did not on the QMFPS after the first semester of a junior/senior level quantum mechanics course. In addition, the performance of graduate students on QMFPS after instruction in the first semester of a core graduate-level quantum mechanics course was significantly better than the performance of undergraduate students at the end of the first semester of an undergraduate quantum mechanics course. However, both undergraduate and graduate students struggled with many questions on the QMFPS. A comparison with the base line data on the validated QMFPS presented here can aid instructors in assessing the effectiveness of their instructional approaches and help them identify the difficulties their students have with quantum formalism and postulates in order to help students develop a solid grasp of the formalism and postulates of quantum mechanics.

## I.    INTRODUCTION

Learning quantum mechanics (QM) is challenging partly because it is abstract as well as nonintuitive and students often transfer ideas from classical mechanics to quantum mechanics inappropriately [1-4]. Several studies have focused on student difficulties with concepts [5-15] and formalism [16-24] in QM. We must help students develop a coherent knowledge structure of the foundational concepts related to the formalism and postulates of quantum mechanics before they can solve novel, complex problems. Furthermore, developing a robust understanding of quantum mechanics requires a solid grasp of linear algebra, differential equations, and special functions [25]. Regardless of the mathematical complexity of quantum mechanics problems, students must develop a functional understanding of quantum mechanics. This entails developing a good knowledge structure of the underlying concepts and be able to reason systematically about relevant quantum mechanics concepts involved while also developing quantitative skills to solve the problems instead of using a plug-and-chug approach.

Research-based conceptual surveys (both free response and multiple-choice format) are useful tools for evaluating student understanding of various topics without focusing heavily on their mathematical skills. Furthermore, carefully developed and validated surveys can play an important role in measuring the effectiveness of a curriculum and instruction. If well-designed multiple-choice pretests and posttests are administered before and after instruction in relevant concepts, they can provide one objective means to measure the effectiveness of a curriculum and instructional approach in a particular course. When compared to free response, multiple choice is free of grader bias and such tests can be graded with great efficiency. Furthermore, the results are objective and amenable to statistical analysis so that different instructional methods or different student populations can be compared. Also, good instructional design requires taking into account the prior knowledge of the students. An effective way to assess the prior knowledge of students, i.e., what the students know before instruction in a particular course, is to administer conceptual surveys as pre-tests. When pre-tests are compared with post-tests, the comparison can give us one objective measure of the effectiveness of instruction.

Several multiple-choice conceptual surveys have been developed for use in physics and astronomy courses [26]. For example, in introductory physics, researchers have developed many multiple-choice surveys to determine the knowledge states of students at the beginning and end of instruction in a particular course and/or topic, e.g., the Force Concept Inventory, Conceptual Survey of Electricity and Magnetism, Rotational and Rolling Motion Survey, Energy and Momentum Survey, etc. [27-30]. In addition, conceptual surveys have been developed for use in QM (quantum mechanics) courses [31-33]. For example, the quantum mechanics conceptual survey (QMCS) [31] was developed for sophomore-level modern physics courses. It focuses on wave functions and probability, wave particle duality, the Schrodinger equation, quantization of states, uncertainty principle, superposition, operators and observables, and tunneling. It contains 12 questions. The quantum mechanics concept assessment (QMCA) [32] is a 31-item survey that focuses on the time-independent Schrodinger equation, time evolution, wave functions and boundary conditions, and probability and it can be used in an upper-division junior/senior level QM course. The quantum mechanics visualization instrument (QMVI) was developed to evaluate students' conceptual understanding of core topics in quantum mechanics in the undergraduate curriculum, especially their visualization skills [33]. Furthermore, the quantum mechanics assessment tool (QMAT) gauges student learning in a first semester junior-level quantum mechanics course and focuses on wave functions, measurement, time dependence, probability, infinite square well, one-dimensional tunneling, and energy levels [34]. The introductory quantum physics conceptual survey (IQPCS) focuses on basic quantum concepts related to quantization and uncertainty [35]. The quantum mechanics survey (QMS) [36] covers topics in non-relativistic quantum mechanics in one spatial dimension typically covered in the first semester of an upper-level undergraduate course and involves concepts such as possible wave functions, bound/scattering states, measurement, expectation values, time dependence of wave function and expectation values, stationary and non-stationary states, role of the Hamiltonian, uncertainty principle, and Ehrenfest's theorem. It can be used in most junior/senior-level quantum mechanics courses if relevant concepts are covered.

However, previously developed conceptual surveys for use in QM courses do not focus explicitly on the postulates or formalism of quantum mechanics. For example, Dirac notation, the Hilbert space, state vectors, physical observables and their corresponding Hermitian operators, compatible and incompatible observables, projection operators and writing operators in terms of their eigenstates and eigenvalues are not covered in other QM conceptual surveys. In addition, other previously developed QM surveys do not include concepts related to spin angular momenta. Therefore, we developed and validated a QM conceptual survey that focuses on the formalism and postulates of quantum mechanics that includes these concepts. Here, we discuss the development and validation of the Quantum Mechanics Formalism and Postulates Survey (QMFPS), which is a 34-item multiple-choice test appropriate for use in an upper-level undergraduate quantum mechanics course as a post-test (after instruction in relevant concepts) or graduate level quantum mechanics course as a pre-test (at the beginning of the course) or post-test [19]. The survey can be used to identify upper-level undergraduate students' final and graduate students' initial and final knowledge states related to the formalism and postulates of quantum mechanics at the beginning and end of a course to assess the effectiveness of a quantum mechanics curriculum in which relevant concepts are covered. The results of the survey can also be used to guide the development of instructional strategies to help students learn these concepts better.

## II. QMFPS SURVEY DEVELOPMENT AND ADMINISTRATION

According to the standards for multiple-choice test design, a high-quality test has five characteristics: reliability, validity, discrimination, good comparative data, and is tailored to the population [37-39]. Furthermore, the development of a well-designed multiple-choice test is an iterative process that involves recognizing the need for the test, formulating the test objectives, constructing test items, performing content validity and reliability check, and distribution [37-39]. Below, we describe the development of the QMFPS and how we ensured that the test was developed based upon the standards of multiple-choice test design.

**Development of the survey**: We recognized the need for a conceptual survey focused on the formalism and postulates of quantum mechanics in that previously developed QM conceptual surveys do not focus explicitly on the postulates or formalism of quantum mechanics. In particular, there are no QM surveys that focus explicitly on Dirac notation, the Hilbert space, state vectors, physical observables and their corresponding Hermitian operators, compatible and incompatible observables, projection operators and writing operators in terms of their eigenstates and eigenvalues are not covered in other QM conceptual surveys. Furthermore, other QM surveys do not include concepts related to spin angular momenta. Therefore, we developed the QMFPS, which focuses on assessing students' conceptual understanding of the formalism and postulates of QM, including Dirac notation, the Hilbert space, state vectors, physical observables and their corresponding Hermitian operators, compatible and incompatible observables, quantum measurement, time-dependence of quantum states and expectation values, and spin angular momenta. The final version of the survey is included on PhysPort [41]. Table I shows one possible categorization of the questions on the survey based upon the concepts, although the categorization may be done in many other ways.

While designing the survey, we focused on making sure that it is valid and reliable [37-39]. Validity refers to the extent to which the test consistently measures whatever it is supposed to measure, and reliability refers to the extent to which the test measures what it measures consistently [37-39]. To ensure that the survey is valid for low-stakes group assessment of QM curriculum and instructional approaches that focus on relevant topics, we consulted with 6 faculty members regarding the goals of their QM courses and topics their students should have learned related to the formalism and postulates of quantum mechanics in upper-level undergraduate QM. In addition to carefully looking through the coverage of these topics in several upper-level undergraduate quantum mechanics classes, we also browsed over several homework, quiz and exam problems that faculty in these courses at the University of Pittsburgh (Pitt) had given to their students in the past when we started the development of the survey. We also gave open-ended questions on relevant topics to students in upper-level QM and interviewed some students one-on-one to get an in-depth understanding of their reasoning behind their responses. These interactions with faculty members and students helped us formulate the test objectives and construct the preliminary test items in initial versions of the survey. We note that the faculty members were not only consulted initially before the development of the survey questions, but we also iterated different versions of the survey with several instructors at Pitt at various stages of the development to ensure that the test content was valid, i.e., that the test items matched the objectives of the test and the test items were accurate, formatted correctly, and were grammatically correct. The faculty members reviewed different versions of the survey several times to examine its appropriateness and relevance for the upper-level quantum mechanics courses and to detect any possible ambiguity in item wording. These valuable comments and feedback from faculty members also helped to ensure that the test was designed with the target population (upper-level undergraduate and graduate students) in mind, i.e., the difficulty level of the questions were appropriate for this target population. In addition to the analysis of the responses to the open-ended questions and interviews, the alternative choices for the multiple-choice questions were informed by prior research on common student difficulties in QM on these topics [1-4, 14-16, 19-24].

**Table I.** One possible categorization of the survey questions, the number of questions that fall in each category, and the question numbers belonging to each category. The number of questions in different categories do not add up to 34 because some questions fall into more than one category.

| Topic | Item number |
|---|---|
| Quantum states | 1,4,7,11,12,13,14,15,17,19 |
| Eigenstates of operators corresponding to physical observables | 1,4,7,14,15,17,18, 20 |
| Time development of quantum states | 3,4,5,6,7,26,32,34 |
| Measurement | 2,3,4,5,7,8,9,13,19,21,23,24,25,27,28,31,32,33,34 |
| Expectation value of observables | 5,10,22 |
| Time dependence of expectation value of observables | 15,16,17,29,30 |
| Commutators/compatibility | 16,17,19,20,27,28 |
| Spin angular momentum | 20,21,22,23,24,25,26,27,28,29,30 |
| Dirac notation | 4,9,10,11,12,13,14,18 |
| Dimensionality of the Hilbert space | 1 |

The individual interviews were conducted with 23 students using a think-aloud protocol [40] at various phases of the test development to better understand students' reasoning processes while they answered the open-ended and multiple-choice questions. Within this interview protocol, students were asked to talk aloud while they answered the questions so that the interviewer could understand their thought processes. The interviews were invaluable and often revealed unnoticed difficulties (not necessarily clear from written responses), and these were incorporated into new versions of the survey. This allowed us to refine the survey further to ensure that the questions were relevant and clearly worded. The interviews also allowed us to further confirm that the difficulty level of the test was appropriate for upper-level undergraduate and graduate students (i.e., the test was designed for the target population in mind).

The final version of the survey can be accessed via the link in ref. [41]. Each question has one correct choice and four incorrect choices. We find that almost all of the students are able to complete the QMFPS in one 50 minute class period after instruction in relevant concepts. Students can answer the QMFPS questions without performing complex calculations, although they do need to understand the basics of linear algebra since that is central to the formalism and postulates of QM. The survey can be used in a junior/senior level undergraduate quantum mechanics course (e.g., at the level of the first four chapters in D. Griffiths' QM textbook [42]) as a post-test, as long as students have learned Dirac notation. It can also be administered in a graduate-level QM course as both a pre-test or post-test to determine students' initial and final knowledge states in regards to the formalism and postulates of QM. While the QMFPS should not be administered as a high-stakes test and the data should be interpreted for the class as a whole to gauge the effectiveness of instruction, it is suggested that students receive some credit for completing the survey in order for students to take it seriously. For example, if the survey is given as a posttest, it can count as a graded low-stakes quiz. If the survey is given as a pretest in a graduate course, it can count as a quiz for which students should be given full credit for trying their best.

**Administration of the validated survey:** The reliability check is performed during a large-scale administration of the final form of the test [37-39]. The validated QMFPS was administered to 464 students from 6 institutions over a period of four years.* Of the 464 students, 350 were undergraduate students and 114 were graduate students. The undergraduate students were enrolled in the first semester of a QM course at the junior/senior level. The graduate students were enrolled in a graduate-level core QM course. The undergraduate students completed the survey as a posttest at the end of their first semester in QM, and the graduate level students completed the survey after at least two months into the first semester of graduate level quantum mechanics. Both the undergraduate and graduate students worked through the survey during a 50-minute class period. Some of the undergraduate students were enrolled in QM courses that used research-based learning tools such as concept tests and quantum interactive learning tutorials ($N = 43$). The survey was given to a subset of these students twice, once at the end of the first semester and then again at the beginning of the second semester after the winter break ($N = 15$). This large-scale administration allowed us to collect comparative data by administrating the test to various groups of students for whom it was designed.

**General Test Statistics:** The average score on the survey after instruction is 41% (including only the first score of the students who took the survey twice). The standard deviation is 20%, with the highest score being 100%. The average score of undergraduate students is 37% and the average score of graduate students is 52%. The fact that the graduate students' performance is better than undergraduate students' performance provides another measure of content validity since graduate students are supposed to know these concepts better overall. There is a significant difference between the graduate and undergraduate students' average scores (p-value of t-test<0.001). Figure 1 shows a histogram of the students' scores on the QMFPS.

The average posttest score for the upper-level students who used concept tests and group discussion and Quantum Interactive Learning Tutorials (QuILTs) was 58% (S.D.=20%). The average posttest score for other undergraduate students who did not use research-based learning tools was 32% (S.D.=16%). There is a significant difference between the scores of students who used research-based learning tools and those who did not (p-value of t-test<0.0001).

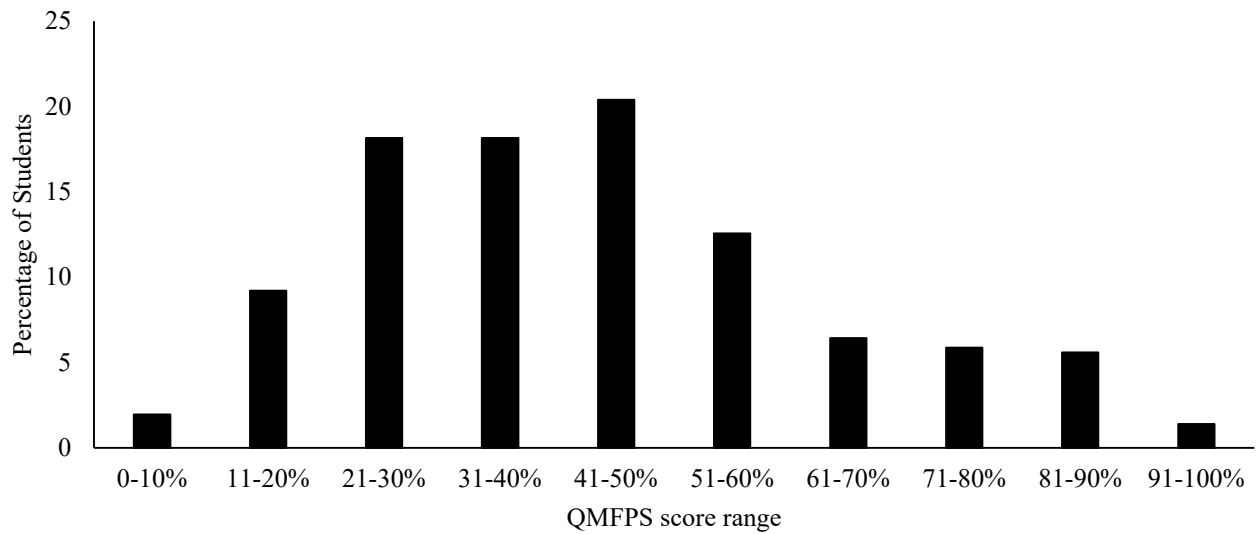

**Figure 1.** Histogram of student scores on the QMFPS.

**Reliability analysis:** We performed various statistical analyses to determine if the QMFPS is reliable. If a test is administered twice at different times to the same sample of students, then one would expect a highly significant correlation between the two test scores (test-retest reliability), assuming that the students' performance is stable and that the test environmental conditions are the same on each occasion [37-39]. Since testing students twice in a very short interval is not practical, one way to determine overall test reliability is via the Kuder-Richardson reliability index (KR-20), which is a measure of the self-consistency of the entire test. According to the standards of test design [37-39], the KR-20 should be higher than 0.7 to ensure that the test is reliable. The KR-20 for the QMFPS is 0.87, indicating that the survey is very reliable.

Performing item analysis can provide further insights into the survey's reliability. Table II shows that, on the QMFPS, the item difficulty (percentage correct) for undergraduate students ranges from 0.13 to 0.70 and the item difficulty for graduate students ranges from 0.16 to 0.90. Table III shows the distribution of student responses for each survey item. We calculated item discrimination for each item on the survey to ensure that the test is reliable. One way to measure item discrimination is by calculating the point-biserial coefficient. It is a measure of consistency of a single test item with the whole test—it reflects the correlation between a student's score on an individual item and his/her score on the entire test [37-39]. The point-biserial coefficient has a possible range of −1 to +1. If an item has a high point-biserial coefficient, then students with high total scores are more likely to answer the item correctly than students with low total scores. A negative point-biserial value indicates that students with low overall scores were more likely to get a particular item correct than those with a high overall score and is an indication that the particular test item is probably defective. Ideally, point-biserial coefficients should be above 0.2 [37-39]. Table II shows the point-biserial coefficients for each item on the QMFPS. The average point-biserial is 0.41 and ranges from 0.20 – 0.62. The standards of test design [34,35] indicate that the survey questions have reasonably good item discrimination. The question with the lowest point-biserial coefficient of 0.20 (Q 32) was also the most difficult question on the test for all students (item difficulty is 0.16).

Another aspect of survey reliability is construct-related validity, which is associated with understanding the nature of the characteristics being measured and the consequences of the uses and interpretations of the results [37-39]. A construct is an individual characteristic that is used to explain the performance on an assessment. For example, mathematical reasoning is a construct that can be used to explain students' performance on a mathematics assessment [37-39]. In our survey, understanding of the formalism and postulates of QM is a construct that can be used to explain performance on the QMFPS. One way to collect evidence of construct validity involves related measures studies. Related measures studies investigate correlations between different assessment measures. For example, one would expect a positive correlation between the Force Concept Inventory and the Force and Motion Conceptual Evaluation since they were designed to measure similar constructs (i.e., students' understanding of force and motion) [37-39]. Therefore, we examined whether students' QMFPS scores were correlated with other validated QM surveys and their performance in quantum mechanics courses to ensure construct validity of the QMFPS. Eighty-two undergraduate students enrolled in the first semester of an undergraduate upper-division QM course were given both the QMFPS and the Quantum Mechanics Survey (QMS), which is a previously validated survey that focuses on students' understanding of non-relativistic QM in one-dimension, after traditional instruction in relevant concepts. Figure 2 shows that there is a strong correlation between students' scores on the QMS and the QMFPS. This correlation provides construct-validity to the QMFPS survey because students who performed well on the QMS are generally likely to have a better foundation in formalism and postulates of quantum mechanics and perform better on the QMFPS. The QMFPS tends to be more difficult for students than the QMS, possibly because it covers more advanced topics as opposed to the QMS which covers QM in one spatial dimension.

**Table II.** Item difficulty (percentage of students answering the question correctly) for undergraduate students (UG), graduate students (G) and all students combined (ALL) and item discrimination (point-biserial coefficient) for each item on the QMFPS.

| Item Number | Item difficulty | | | Item discrimination (point-biserial coefficient) |
|---|---|---|---|---|
| | UG | G | ALL | |
| 1 | 30% | 53% | 37% | 0.49 |
| 2 | 65% | 59% | 63% | 0.38 |
| 3 | 19% | 43% | 26% | 0.59 |
| 4 | 62% | 90% | 70% | 0.48 |
| 5 | 40% | 47% | 42% | 0.31 |
| 6 | 37% | 52% | 42% | 0.37 |
| 7 | 35% | 47% | 38% | 0.33 |
| 8 | 22% | 26% | 23% | 0.35 |
| 9 | 32% | 50% | 37% | 0.53 |
| 10 | 55% | 82% | 63% | 0.53 |
| 11 | 44% | 77% | 53% | 0.49 |
| 12 | 51% | 80% | 59% | 0.50 |
| 13 | 40% | 80% | 51% | 0.55 |
| 14 | 37% | 71% | 47% | 0.42 |
| 15 | 23% | 59% | 39% | 0.42 |
| 16 | 32% | 36% | 33% | 0.23 |
| 17 | 34% | 61% | 42% | 0.42 |
| 18 | 37% | 72% | 48% | 0.44 |
| 19 | 32% | 49% | 37% | 0.26 |
| 20 | 29% | 55% | 37% | 0.43 |
| 21 | 56% | 74% | 61% | 0.48 |
| 22 | 23% | 32% | 26% | 0.34 |
| 23 | 70% | 82% | 73% | 0.38 |
| 24 | 42% | 56% | 46% | 0.37 |
| 25 | 33% | 38% | 34% | 0.42 |
| 26 | 47% | 74% | 55% | 0.50 |
| 27 | 52% | 53% | 53% | 0.27 |
| 28 | 41% | 48% | 43% | 0.34 |
| 29 | 18% | 24% | 19% | 0.39 |
| 30 | 18% | 35% | 22% | 0.39 |
| 31 | 19% | 45% | 27% | 0.54 |
| 32 | 16% | 16% | 16% | 0.20 |
| 33 | 13% | 49% | 25% | 0.62 |
| 34 | 19% | 18% | 19% | 0.31 |

**Table III.** Distribution of undergraduate and graduate student responses to individual questions on the QMFPS. Bolded percentages indicate the correct answer.

|          | Undergraduate students |     |     |     |     | Graduate students |     |     |     |     |
|----------|-----|-----|-----|-----|-----|-----|-----|-----|-----|-----|
| Question | A   | B   | C   | D   | E   | A   | B   | C   | D   | E   |
| 1  | **30%** | 8%  | 15% | 8%  | 39% | **53%** | 2%  | 10% | 3%  | 32% |
| 2  | **65%** | 2%  | 13% | 15% | 5%  | **59%** | 1%  | 13% | 22% | 5%  |
| 3  | 9%  | 25% | 2%  | 44% | **19%** | 4%  | 11% | 3%  | 39% | **43%** |
| 4  | **62%** | 9%  | 13% | 6%  | 9%  | **90%** | 3%  | 3%  | 3%  | 1%  |
| 5  | 7%  | 9%  | 9%  | **40%** | 33% | 2%  | 2%  | 8%  | **47%** | 42% |
| 6  | 4%  | 2%  | 29% | **37%** | 27% | 6%  | 4%  | 5%  | **52%** | 33% |
| 7  | 22% | 9%  | 5%  | **35%** | 29% | 20% | 5%  | 4%  | **47%** | 25% |
| 8  | 36% | 21% | **22%** | 9%  | 12% | 38% | 11% | **26%** | 9%  | 16% |
| 9  | 11% | 27% | 25% | 4%  | **32%** | 2%  | 36% | 8%  | 4%  | **50%** |
| 10 | 9%  | 9%  | 4%  | **55%** | 24% | 4%  | 1%  | 2%  | **82%** | 10% |
| 11 | 37% | 1%  | **44%** | 7%  | 11% | 18% | 0%  | **77%** | 3%  | 2%  |
| 12 | 6%  | 26% | **51%** | 12% | 5%  | 2%  | 15% | **80%** | 1%  | 2%  |
| 13 | 7%  | 10% | 25% | 17% | **40%** | 0%  | 3%  | 8%  | 9%  | **80%** |
| 14 | **37%** | 12% | 28% | 13% | 9%  | **71%** | 1%  | 20% | 6%  | 2%  |
| 15 | 14% | **30%** | 19% | 13% | 23% | 7%  | **59%** | 8%  | 13% | 12% |
| 16 | 10% | 19% | 17% | 21% | **32%** | 13% | 6%  | 20% | 26% | **36%** |
| 17 | **34%** | 12% | 6%  | 22% | 25% | **61%** | 6%  | 3%  | 14% | 16% |
| 18 | 16% | **37%** | 25% | 5%  | 15% | 3%  | **72%** | 16% | 1%  | 8%  |
| 19 | 9%  | 23% | 28% | 7%  | **32%** | 11% | 8%  | 27% | 6%  | **49%** |
| 20 | 24% | **29%** | 11% | 9%  | 25% | 24% | **55%** | 2%  | 2%  | 16% |
| 21 | 2%  | 22% | 7%  | **56%** | 11% | 0%  | 4%  | 1%  | **74%** | 20% |
| 22 | 11% | 18% | 21% | 25% | **23%** | 14% | 18% | 10% | 25% | **32%** |
| 23 | 18% | **70%** | 5%  | 3%  | 3%  | 13% | **82%** | 3%  | 0%  | 2%  |
| 24 | **42%** | 12% | 12% | 25% | 5%  | **56%** | 10% | 13% | 19% | 2%  |
| 25 | 21% | 11% | 17% | **33%** | 13% | 22% | 11% | 17% | **38%** | 12% |
| 26 | 12% | 27% | 7%  | **47%** | 3%  | 5%  | 13% | 1%  | **74%** | 6%  |
| 27 | 28% | 5%  | **52%** | 6%  | 5%  | 37% | 1%  | **53%** | 0%  | 5%  |
| 28 | 12% | **41%** | 16% | 19% | 6%  | 4%  | **48%** | 11% | 26% | 9%  |
| 29 | 11% | 4%  | **18%** | 46% | 15% | 10% | 4%  | **24%** | 35% | 26% |
| 30 | **18%** | 4%  | 9%  | 51% | 12% | **35%** | 2%  | 6%  | 45% | 10% |
| 31 | 13% | 10% | **19%** | 25% | 27% | 9%  | 14% | **45%** | 13% | 18% |
| 32 | 37% | 15% | **16%** | 11% | 15% | 55% | 13% | **16%** | 7%  | 5%  |
| 33 | 15% | 11% | **13%** | 26% | 28% | 9%  | 10% | **49%** | 11% | 16% |
| 34 | 38% | 13% | **19%** | 9%  | 15% | 53% | 14% | **18%** | 5%  | 5%  |

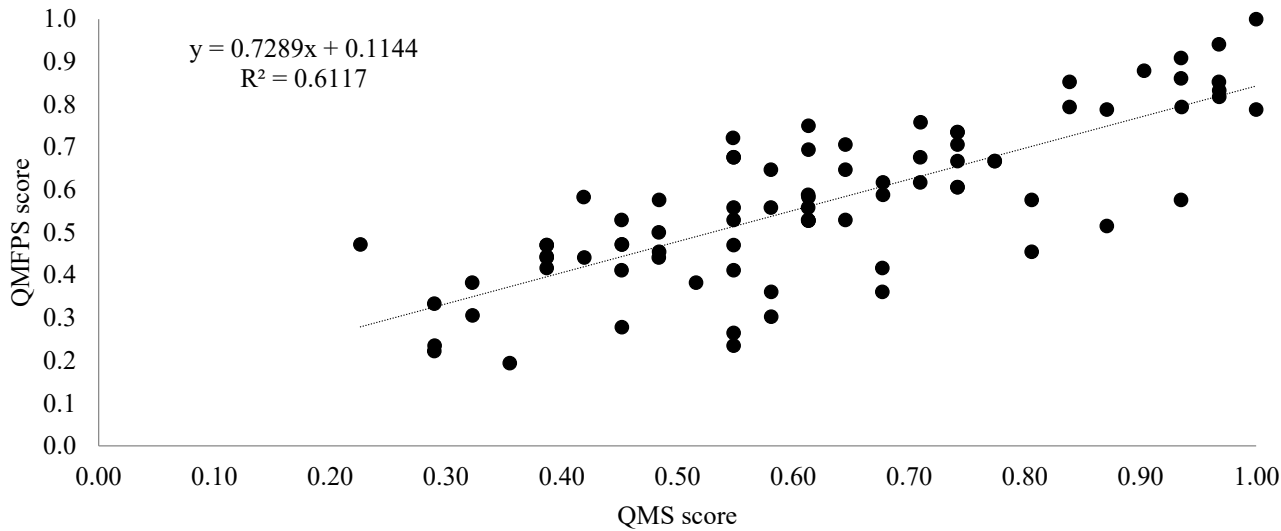

**Figure 2**. Correlation between 82 students' QMFPS scores and their QMS scores. The coefficient of determination is $R^2$ and the correlation coefficient is R=0.78.

In addition, 44 graduate students enrolled in the first semester of a graduate-level core quantum mechanics course were given the QMFPS at the end of the semester. Figure 3 shows that there is a moderate correlation between students' scores on the QMFPS and their final exam in the graduate-level QM course. This correlation provides further evidence of construct validity of the QMFPS since the concepts covered in the final exam were similar to those covered in the QMFPS.

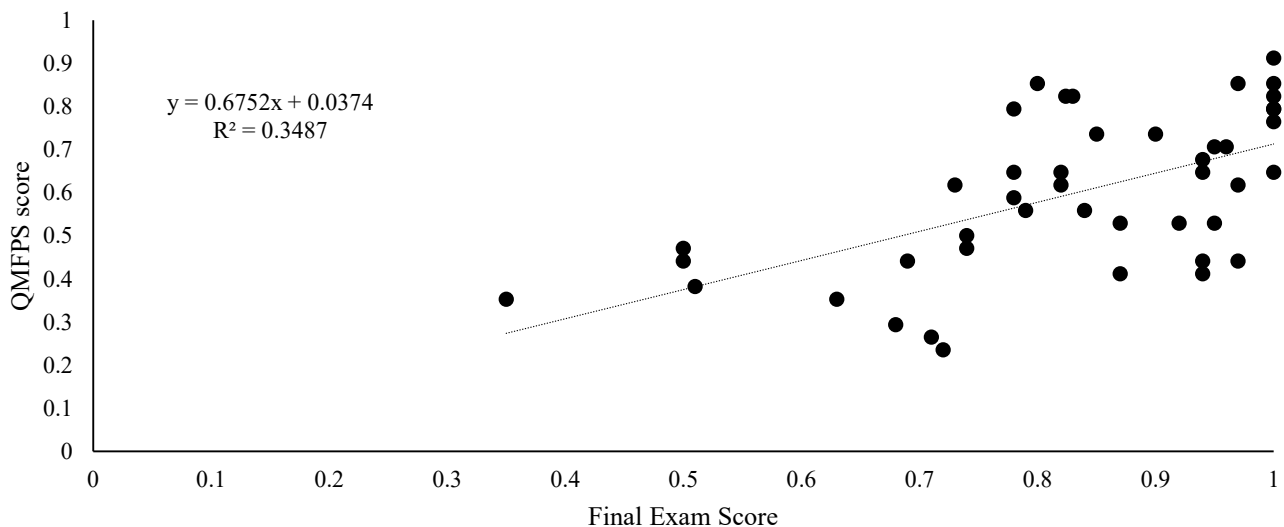

**Figure 3**. Correlation between 44 graduate students' QMFPS scores and their final exam score in a core graduate level quantum mechanics course. The coefficient of determination is $R^2$ and the correlation coefficient is R=0.58.

## III. SUMMARY

Learning QM is challenging for students partly because of the "paradigm shift" from classical mechanics to quantum mechanics as well as the mathematical expertise required to solve problems. Students in traditionally taught and evaluated QM courses may be able to "hide" their lack of conceptual understanding of the formalism and postulates of QM behind their mathematical skills [3]. However, in order for students to develop functional understanding, it is important to close the gap [43] between conceptual and quantitative problem-solving by assessing both types of learning. We have developed a conceptual survey that assesses students' conceptual knowledge of the formalism and postulates of QM, which are topics that instructors of QM courses agree are important to cover [25]. The development of the test followed the standards of multiple-choice test design [37-39], and we ensured that the test was valid and reliable, had good discrimination, was tailored to the population, and that we collected good comparative data [37-39].

Details of student difficulties found via QMFPS is beyond the scope of this paper and will be discussed elsewhere. Student responses to questions on the QMFPS can be used as a formative assessment [44,45] to help instructors identify common student difficulties and guide the design of instructional strategies and learning tools to improve students' understanding. This survey can also be administered to students after instruction in the relevant concepts to evaluate the effectiveness of instruction on relevant topics in a particular course.

Furthermore, we found that students who were enrolled in QM courses that used active learning methods, such as peer instruction and tutorials, performed better on the QMFPS than those who did not. These approaches included active learning techniques [46-85] such as peer instruction [52,53], tutorials [54-68], cooperative group problem solving [69], and Just-In-Time-Teaching [70,71], to help students develop a coherent knowledge structure of the formalism and postulates of QM. In addition, we found that, although graduate students performed significantly better than undergraduate students, their average overall score was not very high. This may partly be due to the fact that graduate students who were taught primarily via traditional approaches may have developed algorithmic skills to solve problems on their exams, which often reward "plug and chug" approaches, but lack a conceptual understanding of quantum mechanics. However, even graduate students may not be motivated to develop a coherent knowledge structure of QM if course assessments only focus on quantitative reasoning. Therefore, in order to help students develop a functional knowledge of quantum mechanics, we suggest that the learning goals for upper-level QM include proficiency in concepts covered in the QMFPS and emphasize the connection between conceptual understanding and mathematical formalism. Furthermore, instructors of graduate-level QM courses can reflect on their students' responses on the QMFPS to design instruction that helps to "close the gap" between students' conceptual understanding and quantitative problem solving.

## ACKNOWLEDGEMENTS

The authors are grateful to the faculty members who reviewed and provided feedback on the survey at several stages. We are also thankful to the students who participated in interviews which greatly helped in the design of the survey. We also thank the various faculty members who gave feedback on the survey and administered this survey in their quantum mechanics classes. We are especially grateful to R.P. Devaty for his feedback during the development of the QMFPS and writing of this manuscript. This work is supported by the National Science Foundation grant number NSF-1806691.

# REFERENCES

*Some instructors did not administer the last four questions of the QMFPS. There were 358 students who answered the first thirty questions on the QMFPS.

1. C. Singh, Student understanding of quantum mechanics, Am. J. Phys. **69**, 885 (2001).
2. C. Singh, Student understanding of quantum mechanics at the beginning of graduate instruction, Am. J. Phys. **76**, 277 (2008).
3. E. Marshman and C. Singh, Framework for understanding student difficulties in quantum mechanics, Phys. Rev. ST Phys. Educ. Res. **11**, 020117 (2015).
4. C. Singh and E. Marshman, Review of student difficulties in quantum mechanics, Phys. Rev. Special Topics Phys. Educ. Res. **11**, 020119 (2015).
5. M. Wittmann, R. Steinberg and E. Redish, Investigating student understanding of quantum physics: Spontaneous models of conductivity, Am. J. Phys. **70,** 218 (2002).
6. M. Wittmann, J. Morgan, and L. Bao, Addressing student models of energy loss in quantum tunneling, Eur. J. Phys. **26**, 939 (2005).
7. P. Jolly, D. Zollman, S. Rebello and A. Dimitrova, Visualizing potential energy diagrams, Am. J. Phys. **66**, 57 (1998).
8. P.C. Garcia Quijas and L.M. Arevala Aguilar, Overcoming misconceptions in quantum mechanics with the time evolution operator, Eur. J. Phys. **28**, 147 (2007).
9. S. Sharma and P.K. Ahluwalia, Diagnosing alternative conceptions of Fermi energy among undergraduate students, Eur. J. Phys. **33**, 883 (2012).
10. D. Domert, C. Linder and A. Ingerman, Probability as a conceptual hurdle to understanding one-dimensional quantum scattering and tunneling, Eur. J. Phys. **26**, 47 (2005).
11. C. Singh, Transfer of learning in quantum mechanics, in *Proc. Phys. Educ. Res. Conf.,* AIP Conf. Proc. Melville NY **790,** p. 23 (2005) https://doi.org/10.1063/1.2084692.
12. C. Singh and G. Zhu, Cognitive issues in learning advanced physics: An example from quantum mechanics, in *Proc. Phys. Educ. Res. Conf.* (2009) https://doi.org/10.1063/1.3266755.
13. C. Singh, Assessing and improving student understanding of quantum mechanics, AIP Conf. Proc. Melville NY **818**, 69–72 (2006).
14. G. Zhu and C. Singh, Improving students' understanding of quantum measurement: I. Investigation of difficulties, Phys. Rev. Special Topics Phys. Educ. Res. **8**, 010117 (2012).
15. S. Lin and C. Singh, Categorization of quantum mechanics problems by professors and students, Eur. J. Phys. **31**, 57 (2010).
16. C. Singh and E. Marshman, Analogous patterns of student reasoning difficulties in introductory physics and upper-level quantum mechanics, in *Proc. Phys. Educ. Res. Conf.* (2014) https://doi.org/10.1119/perc.2013.inv.010.
17. E. Marshman, C. Keebaugh and C. Singh, Investigating student difficulties with the corrections to the energy of the hydrogen atom for the strong and weak field Zeeman effect, *Proc. Physics Education Research Conf.* (2018) https://doi.org/10.1119/perc.2017.pr.060.
18. C. Singh, E. Marshman and C. Keebaugh, Student difficulties with finding the fine structure corrections to the energy spectrum of the hydrogen atom using degenerate perturbation theory, Proc. Phys. Educ. Res. Conf. (2018) https://doi.org/10.1119/perc.2017.pr.088
19. E. Marshman, Doctoral Dissertation, accessed at http://d-scholarship.pitt.edu/25547/.

20. E. Marshman and C. Singh, Student difficulties with quantum states while translating state vectors in Dirac notation to wave functions in position and momentum representations, in *Proc. Phys. Educ. Res. Conf.* (2015) https://doi.org/10.1119/perc.2015.pr.048.
21. C. Singh, and E. Marshman, Student difficulties with determining expectation values in quantum mechanics, in *Proc. Physics Education Research Conf.* (2016) https://doi.org/10.1119/perc.2016.pr.075.
22. E. Marshman, and C. Singh, Student difficulties with representations of quantum operators corresponding to observables, in *Proc. Phys. Educ. Res. Conf.* (2016) https://doi.org/10.1119/perc.2016.pr.049.
23. C. Singh and E. Marshman, Investigating student difficulties with Dirac Notation, in *Proc. Physics Education Research Conf.* (2014) https://doi.org/10.1119/perc.2013.pr.074.
24. E. Marshman and C. Singh, Investigating student difficulties with time-dependence of expectation values in quantum mechanics, in *Proc. Phys. Educ. Res. Conf.* (2014) https://doi.org/10.1119/perc.2013.pr.049.
25. S. Siddiqui and C. Singh, How diverse are physics instructors' attitudes and approaches to teaching undergraduate-level quantum mechanics? Eur. J. Phys. **38,** 035703 (2017).
26. A. Madsen, S. McKagan, and E. Sayre, Resource Letter RBAI-1: Research-based assessment instruments in physics and astronomy, Am. J. Phys. **85**, 2017.
27. D. Hestenes, M. Wells, G. Swackhamer, Force concept inventory, The Phys. Teach. **30**, 141 (1992).
28. D.P. Maloney, T.L. O'Kuma, C.J. Hieggelke, and A.V. Heuvelen, Surveying students' conceptual knowledge of electricity and magnetism, Am. J. Phys. **69**, S12 (2001).
29. L. Rimoldini and C. Singh, Student understanding of rotational and rolling motion concepts, Phys. Rev. Special Topics Phys. Educ. Res., **1**, 010102 (2005).
30. C. Singh and D. Rosengrant, Multiple-choice test of energy and momentum concepts, Am. J. Phys. **71**, 607 (2003).
31. S. B. McKagan et al., Design and validation of the quantum mechanics conceptual survey, Phys. Rev. ST Phys. Educ. Res. **6**, 020121 (2010).
32. H. Sadaghiani and S. Pollock, Quantum mechanics concept assessment: Development and validation study, Phys. Rev. ST Phys. Educ. Res. **11**, 010110 (2014).
33. R. Robinett and E. Cataloglu, Testing the development of student conceptual and visualization understanding in quantum mechanics through the undergraduate career, Am. J. Phys. **70**, 238 (2002).
34. S. Goldhaber, S. Pollock, M. Dubson, P. Beale, and K. Perkins, *Transforming Upper Division Quantum Mechanics: Learning Goals and Assessment* 2009 Physics Education Research Conference, *AIP Conf. Proc* **1179** 145-148 (2009).
35. S. Wuttiprom, M.D. Sharma, I.D. Johnston, R. Chitaree and C. Soankwan, Development and use of a conceptual survey in introductory quantum physics, Intl. J. Sci. Ed. **31**(5), 631-654 (2009).
36. G. Zhu and C. Singh, Surveying students' understanding of quantum mechanics in one spatial dimension, Am. J. Phys. **80**, 252-259 (2012).
37. G. Aubrecht and J. Aubrecht, Constructing objective tests, Am. J. Phys. **51**, 613–620 (1983) and A. J. Nitko, *Educational Assessments of Students* (Prentice-Hall/Merrill, Englewood Cliffs, NJ, 1996).
38. P.V. Engelhardt, An Introduction to Classical Test Theory as Applied to Conceptual Multiple-choice Tests, in *Getting Started in PER*, ed. C. Henderson and K. Harper (AAPT, College Park, MD, 2009).
39. W. Adams and C. Wieman, Development and validation of instruments to measure learning of expert-like thinking, International Journal of Science Education **33**(9), 1289-1312 (2011).

40. M. Chi, Thinking aloud, in *The Think Aloud Method: A Practical Guide to Modeling Cognitive Processes*, edited by M. W. Van Someren, Y. F. Barnard, and J. A. C. Sandberg (Academic Press, London, 1994).
41. https://www.appendix.org/assessments/assessment.cfm?A=QMFPS
42. D. J. Griffiths, *Introduction to Quantum Mechanics* (Prentice Hall, Upper Saddle River, NJ, 1995).
43. C. Singh, Closing the gap between teaching and assessment, written for the report of the Association of American Universities (AAU) 2014, accessed at http://rescorp.org/gdresources/publications/effectivebook.pdf.
44. National Research Council 2001, *Knowing What Students Know: The Science and Design of Educational Assessment. Committee on the Foundations of Assessment*, edited by J. Pellegrino, N. Chudwosky, and R. Glaser (National Academy Press, Washington, DC, 2001).
45. P. Black and D. Wiliam, Assessment and classroom learning, Assessment in Education **5** (1), 7 (1998); P. Black and D. William, Inside the black box: Raising standards through classroom assessment, Phi Delta Kappan **92** (1), 81 (2010); P. Black and D. William, Developing the theory of formative assessment, Educational Assessment, Evaluation and Accountability **21** (1), 5 (2009); P. Black, C. Harrison, C. Lee, B. Marshall, and D. William, *Assessment for Learning: Putting It Into Practice* (Open University Press, Buckingham, 2003).
46. D. Zollman, N. Rebello and K. Hogg, Quantum mechanics for everyone: Hands-on activities integrated with technology, Am. J. Phys. **70**, 252 (2002).
47. A. Kohnle et al., A new introductory quantum mechanics curriculum, Eur. J. Phys. **35**, 015001 (2014).
48. A. Kohnle et al., Developing and evaluating animations for teaching quantum mechanics concepts, Eur. J. Phys. **31**, 1441 (2010).
49. R. Muller and H. Wiesner, Teaching quantum mechanics on an introductory level, Am. J. Phys. **70**, 200 (2002).
50. C. Singh, M. Belloni, and W. Christian, Improving student's understanding of quantum mechanics, Phys. Today **8**(1), 43–49 (2006).
51. B. Brown, C. Singh and A. J. Mason, The effect of giving explicit incentives to correct mistakes on subsequent problem solving in quantum mechanics, in *Proc. Physics Education Research Conf.* (2015) https://doi.org/10.1119/perc.2015.pr.012.
52. E. Mazur, *Peer Instruction: A User's Manual* (Prentice Hall, Upper Saddle River, NJ, 1997).
53. C. Singh and G. Zhu, Improving students' understanding of quantum mechanics by using peer instruction tools, in *Proceedings of the 2011 Phys. Ed. Res. Conf.,* AIP Conf. Proc. Melville, New York **1413**, 77-80 (2012) https://doi.org/10.1063/1.3679998.
54. C. Singh, Interactive learning tutorials on quantum mechanics, Am. J. Phys. **76**(4), 400 (2008).
55. G. Zhu and C. Singh, Improving students' understanding of quantum mechanics via the Stern Gerlach experiment, Am. J. Phys. **79**(5), 499 (2011).
56. C. Singh, Helping students learn quantum mechanics for quantum computing, in *Proc. Physics Education Research Conf.,* AIP Conf. Proc. Melville, New York **883**, p. 42 (2007) https://doi.org/10.1063/1.2508687.
57. G. Zhu and C. Singh, Improving students' understanding of quantum measurement: II. Development of research-based learning tools, Phys. Rev. ST PER **8**, 010118 (2012).
58. G. Zhu and C. Singh, Improving student understanding of addition of angular momentum in quantum mechanics, Phys. Rev. ST PER **9**, 010101 (2013).
59. S. DeVore and C. Singh, Development of an interactive tutorial on quantum key distribution, in *Proc. Physics Education Research Conf.* (2015) https://doi.org/10.1119/perc.2014.pr.011.

60. B. Brown B and C. Singh, Development and evaluation of a quantum interactive learning tutorial on Larmor Precession of spin, in *Proc. Physics Education Research Conf.* (2015) https://doi.org/10.1119/perc.2014.pr.008.
61. E. Marshman and C. Singh, Interactive tutorial to improve student understanding of single photon experiments involving a Mach-Zehnder Interferometer, Eur. J. Phys. **37**, 024001 (2016).
62. E. Marshman and C. Singh, Investigating and improving student understanding of quantum mechanics in the context of single photon interference, Phys. Rev. Phys. Educ. Res. **13**, 010117 (2017).
63. E. Marshman and C. Singh, Investigating and improving student understanding of quantum mechanical observables and their corresponding operators in Dirac notation, Eur. J. Phys. **39**, 015707 (2017).
64. E. Marshman and C. Singh, Investigating and improving student understanding of the expectation values of observables in quantum mechanics, Eur. J. Phys. **38** (4), 045701 (2017).
65. E. Marshman and C. Singh, Investigating and improving student understanding of the probability distributions for measuring physical observables in quantum mechanics, Eur. J. Phys. **38** (2), 025705 (2017).
66. A. Maries, R. Sayer and C. Singh, Effectiveness of interactive tutorials in promoting "which-path" information reasoning in advanced quantum mechanics, Phys. Rev. PER **13**, 020115 (2017).
67. R. Sayer, A. Maries and C. Singh, A quantum interactive learning tutorial on the double-slit experiment to improve student understanding of quantum mechanics, Phys. Rev. PER **13,** 010123 (2017).
68. C. Keebaugh, E. Marshman, and C. Singh, Developing and evaluating an interactive tutorial on degenerate perturbation theory, in *Proc. Phys. Educ. Res. Conf.* (2017) https://doi.org/10.1119/perc.2016.pr.041.
69. P. Heller, R. Keith, and S. Anderson, Teaching problem solving through cooperative grouping. Part 1: Group versus individual problem solving, Am. J. Phys. **60**, 627 (1992).
70. G. Novak, E. Patterson, A. Gavrin, W. Christian, and K. Forinash, Just in time teaching, Am. J. Phys. **67**, 937 (1999).
71. R. Sayer, E. Marshman and C. Singh, A case study evaluating Just-in-Time Teaching and Peer Instruction using clickers in a quantum mechanics course, Phys. Rev. PER **12** 020133 (2016).
72. C. Singh, Student difficulties with quantum mechanics formalism in *Proc. Phys. Educ. Res. Conf.*, AIP Conf. Proc. Melville NY **883**, p.185 (2007) https://aip.scitation.org/doi/abs/10.1063/1.2508723
73. C. Singh and G. Zhu, Improving students' understanding of quantum mechanics by using peer instruction tools, in *Proc. Phys. Educ. Res. Conf.*, AIP Conf. Proc. Melville NY **1413**, p. 77 (2012), https://doi.org/10.1063/1.3679998.
74. A. J. Mason and C. Singh, Do advanced students learn from their mistakes without explicit intervention? Am. J. Phys. **78**, 760 (2010).
75. B. Brown, A. Mason, and C. Singh, Improving performance in quantum mechanics with explicit incentives to correct mistakes, Phys. Rev. PER **12**, 010121 (2016).
76. C. Keebaugh, E. Marshman and C. Singh, Improving student understanding of corrections to the energy spectrum of the hydrogen atom for the Zeeman effect, Phys. Rev. PER **15**, 010113 (2018).
77. C. Keebaugh, E. Marshman, and C. Singh, Investigating and addressing student difficulties with the corrections to the energies of the hydrogen atom for the strong and weak field Zeeman effect," Eur. J. Phys. **39**, 045701 (2018).

78. C. Keebaugh, E. Marshman, and C. Singh, Investigating and addressing student difficulties with a good basis for finding perturbative corrections in the context of degenerate perturbation theory, Eur. J. Phys. **39**, 055701 (2018).
79. C. Keebaugh, E. Marshman, and C. Singh, Student difficulties with the number of distinct many-particle states for a system of non-interacting identical particles with a fixed number of available single-particle states, in *Proc. Phys. Educ. Res. Conf.* 2018 http://doi.org/10.1119/perc.2018.pr.Keebaugh
80. C. Keebaugh, E. Marshman and C. Singh, Improving student understanding of fine structure corrections to the energy spectrum of the hydrogen atom, Am. J. Phys. **87** (7) 594 (2019) https://aapt.scitation.org/doi/10.1119/1.5110473
81. C. Keebaugh, E. Marshman and C. Singh, Improving student understanding of a system of identical particles with a fixed total energy, Am. J. Phys. **87** (7), 583 (2019) https://aapt.scitation.org/doi/10.1119/1.5109862
82. P. Justice, E. Marshman, and C. Singh, Development and validation of a sequence of clicker questions for helping students learn addition of angular momentum in quantum mechanics, 2018 Phys. Educ. Res. Conf. Proc., (2018) https://doi.org/:10.1119/perc.2018.pr.Justice
83. P. Justice, E. Marshman and C. Singh, Improving student understanding of quantum mechanics underlying the Stern-Gerlach experiment using a research-validated multiple-choice question sequence, Eur. J. Phys. **40**, 055702 (2019).
84. P. Justice, E. Marshman and C. Singh, Student understanding of Fermi energy, the Fermi-Dirac distribution and total electronic energy of a free electron gas, Eur. J. Phys. **41**, 015704 (2020).
85. P. Justice, E. Marshman and C. Singh, Development, validation and in-class evaluation of a sequence of clicker questions on Larmor precession of spin in quantum mechanics, 2019 Phys. Educ. Res. Conf. Proc., (2020) https://doi.org/10.1119/perc.2019.pr.Justice

**Supplemental Material on PhysPort**

**Quantum Mechanics Formalism and Postulates Survey**

Definitions, notation, and instructions:

\* For a spinless particle confined in one spatial dimension, the expectation value of a time-independent physical observable $Q$ in a state $|\Psi(t)\rangle$ at time $t$ in position space is

$\langle Q \rangle = \langle \Psi(t)|\hat{Q}|\Psi(t)\rangle = \int_{-\infty}^{\infty} \Psi^*(x,t) Q\left(x, -i\hbar\frac{\partial}{\partial x}\right) \Psi(x,t) dx$. For the special case $t = 0$, we will write simply $|\Psi\rangle \equiv |\Psi(0)\rangle$ and $\Psi(x) \equiv \Psi(x,0)$.

\* Notation $\left(\hat{Q}|\Psi\rangle\right)^\dagger = \left(|\hat{Q}\Psi\rangle\right)^\dagger = \langle\Psi|\hat{Q}^\dagger = \langle\hat{Q}^\dagger\Psi|$

\* $\hat{x}, \hat{p}, \hat{H}$ are generic symbols for the position, momentum and Hamiltonian operators, respectively, for a given quantum system.

\* $|x\rangle, |p\rangle$, and $|n\rangle$ are eigenstates of position, momentum and Hamiltonian operators with eigenvalues $x$, $p$, and $E_n$, respectively. The eigenvalue equation for a generic hermitian operator $\hat{Q}$ with discrete and continuous eigenvalue spectra is given by $\hat{Q}|q_n\rangle = q_n|q_n\rangle$ and $\hat{Q}|q\rangle = q|q\rangle$, respectively.

* For a particle confined in one spatial dimension, the momentum eigenfunction in position space is $Ae^{ipx/\hbar}$ with eigenvalue $p$, where $A$ is a constant.
* Ignore the normalization issues of the position eigenstates and momentum eigenstates.
* The Schrödinger representation is used throughout; there is NO <u>explicit</u> time dependence to any of the observables in any question in the survey.
* When the summation or integration limits are not given explicitly, the summation/integration is over all possible values of the summation/integration variable.
* All measurements are ideal.
* $\hat{S}_x$, $\hat{S}_y$, and $\hat{S}_z$ are the $x$, $y$, and $z$ components of the spin angular momentum operator (or spin) and $\hat{S}^2$ is the square of the spin angular momentum operator.

**In all questions below, the correct answer is bolded.**

**Questions 1-3 refer to the following system: A particle interacts with a one-dimensional infinite square well of width $a$ ($V(x) = 0$ for $0 \le x \le a$ and $V(x) = +\infty$ otherwise). The stationary state wavefunctions are $\psi_n(x) = \sqrt{\frac{2}{a}} \sin\left(\frac{n\pi x}{a}\right)$ and the allowed energies are $E_n = \frac{n^2 \pi^2 \hbar^2}{2ma^2}$ where $n = 1, 2, 3 \ldots \infty$.**

1. Choose all of the following statements that are correct for a particle interacting with a one dimensional (1D) infinite square well.
   (1) The appropriate Hilbert space for this system is one dimensional.
   (2) The energy eigenstates of the system form a basis in a 1D Hilbert space.
   (3) The position eigenstates of the system form a basis in a 1D Hilbert space.
   **A. none of the above**   B. 1 only   C. 2 only   D. 3 only   E. all of the above

2. The wavefunction at time $t = 0$ is $\sqrt{\frac{2}{5}}\psi_1(x) + \sqrt{\frac{3}{5}}\psi_2(x)$ when you perform a measurement of the

energy. Choose all of the following statements that are correct.

(1) The measurement of the energy will yield either $E_1$ or $E_2$.

(2) The spatial part of the **normalized** wavefunction (excluding the time part) after the energy measurement is either $\sqrt{\frac{2}{5}}\psi_1(x)$ or $\sqrt{\frac{3}{5}}\psi_2(x)$.

(3) The measurement of the energy will yield $\frac{2}{5}E_1 + \frac{3}{5}E_2$.

A. 1 only  B. 2 only  C. 3 only  D. 1 and 2 only  E. 1 and 3 only

3. Consider the following wavefunction at time $t = 0$: $\Psi(x, 0) = Ax(a - x)$ for $0 \le x \le a$, where $A$ is a suitable normalization constant. Which one of the following is the probability density $|\Psi(x, t)|^2$, at time $t > 0$?

A. $|\Psi(x, t)|^2 = |A|^2 x^2 (a - x)^2 \cos^2\left(\frac{Et}{\hbar}\right)$, where $E$ is the expectation value of energy.

B. $|\Psi(x, t)|^2 = |A|^2 x^2 (a - x)^2 e^{\frac{-2iEt}{\hbar}}$, where $E$ is the expectation value of energy.

C. $|\Psi(x, t)|^2 = |A|^2 x^2 (a - x)^2 \sin^2\left(\frac{Et}{\hbar}\right)$, where $E$ is the expectation value of energy.

D. $|\Psi(x, t)|^2 = |A|^2 x^2 (a - x)^2$, which is time-independent.

E. None of the above.

4. Suppose $|\Psi\rangle$ is a generic state and the energy eigenstates $|n\rangle$ are such that $\hat{H}|n\rangle = E_n|n\rangle$, where $n = 1, 2, 3 \ldots \infty$. Choose all of the following statements that are correct.

(1) $|\Psi\rangle = \sum_n \langle n|\Psi\rangle |n\rangle$

(2) $e^{-i\hat{H}t/\hbar}|\Psi\rangle = \sum_n e^{-iE_n t/\hbar}\langle n|\Psi\rangle |n\rangle$

(3) If you measure the energy of the system in the state $|\Psi\rangle$, the probability of obtaining $E_n$ and collapsing the state to $|n\rangle$ is $|\langle n|\Psi\rangle|^2$.

A. all of the above  B. 1 and 2 only  C. 1 and 3 only  D. 2 and 3 only  E. 3 only

5. Suppose $|q_n\rangle$ are the eigenstates of an operator $\hat{Q}$ corresponding to a physical observable with a discrete spectrum of non-degenerate eigenvalues $q_n$ where $n = 1, 2, \ldots \infty$. At time $t = 0$, the state of the system is $|\Psi\rangle = \sum_n c_n(t = 0)|q_n\rangle$. Choose all of the following statements that are necessarily correct.

(1) $\langle Q\rangle = \sum_n |c_n(0)|^2 q_n$ at time $t = 0$.

(2) $c_n(t) = e^{-iq_n t/\hbar} c_n(0)$ at time $t > 0$.

(3) Experimentally, $|c_n(0)|^2$ for various $n$ can be estimated by measuring $Q$ in an ensemble of identically prepared systems in state $|\Psi\rangle$ at time $t = 0$.

A. 1 only  B. 2 only  C. 3 only  **D. 1 and 3 only**  E. all of the above

**Questions 6-7 refer to the following system: A particle with mass $m$ interacts with a one-dimensional simple harmonic oscillator well $V(x) = \frac{Kx^2}{2}$. The stationary states are $\psi_n(x)$ and the allowed energies are $E_n = \left(n + \frac{1}{2}\right)\hbar\omega$, where $n = 0, 1, 2 \ldots \infty$ and $\omega = \sqrt{\frac{K}{m}}$.**

6. $|\Psi(0)\rangle$ is the initial state of the system at time $t = 0$ and $\hat{H}$ is the Hamiltonian operator. Choose all of the following statements that are necessarily correct for all times $t > 0$.

   (1) $e^{-i\hat{H}t/\hbar}$ is a hermitian operator.

   (2) $e^{-i\hat{H}t/\hbar}$ is a unitary operator.

   (3) The state of the system at time $t > 0$ is $|\Psi(t)\rangle = e^{-i\hat{H}t/\hbar}|\Psi(0)\rangle$

A. 2 only  B. 1 and 2 only  C. 1 and 3 only  **D. 2 and 3 only**  E. all of the above

7. Suppose you perform a measurement of the position of the particle when it is in the first excited state of a one dimensional simple harmonic oscillator potential energy well. Choose all of the following statements that are correct about this experiment:

   (1) Right after the position measurement, the wavefunction will be peaked about a particular value of position.

   (2) The wavefunction will not go back to the first excited state wavefunction, even if you wait for a long time after the position measurement.

   (3) A long time after the position measurement, the wavefunction will go back to the first excited state wavefunction.

A. 1 only  B. 2 only  C. 3 only  **D. 1 and 2 only**  E. 1 and 3 only

8. Consider the following conversation between Andy and Caroline about the measurement of energy in a state $|\Psi\rangle$ which is not an energy eigenstate.

**Andy**: When an operator $\hat{H}$ corresponding to energy acts on a generic state $|\Psi\rangle$, it corresponds to a measurement of energy. Therefore, $\hat{H}|\Psi\rangle = E_n|\Psi\rangle$, where $E_n$ is the observed value of energy.

**Caroline**: No. The measurement collapses the state so $\hat{H}|\Psi\rangle = E_n|n\rangle$, where $|\Psi\rangle$ on the left hand side is the original state before the measurement and $|n\rangle$ on the right hand side of the equation is the state in which the system collapses after the measurement and it is an eigenstate of $\hat{H}$ with eigenvalue $E_n$.

With whom do you agree?

A. Agree with Caroline only
B. Agree with Andy only
C. **Agree with neither**
D. Agree with both
E. The answer depends on the details of the state $|\Psi\rangle$, which is a linear superposition of energy eigenstates.

**For a spinless particle confined in one spatial dimension, the state of the quantum system at time $t = 0$ is denoted by $|\Psi\rangle$ in the Hilbert space. $|x\rangle$ and $|p\rangle$ are the eigenstates of position and momentum operators. Answer questions 9 to 20.**

9. An operator $\hat{Q}$ corresponding to a physical observable $Q$ has a continuous non-degenerate spectrum of eigenvalues. The states $\{|q\rangle\}$ are eigenstates of $\hat{Q}$ with eigenvalues $q$. At time $t = 0$, the state of the system is $|\Psi\rangle$. Choose all of the following statements that are correct.

    (1) A measurement of the observable $Q$ must return one of the eigenvalues of the operator $\hat{Q}$.
    (2) If you measure $Q$ at time $t = 0$, the probability of obtaining an outcome between $q$ and $q + dq$ is $|\langle q|\Psi\rangle|^2 dq$.
    (3) If you measure $Q$ at time $t = 0$, the probability of obtaining an outcome between $q$ and $q + dq$ is $\left|\int_{-\infty}^{\infty} \psi_q^*(x)\Psi(x)dx\right|^2 dq$ in which $\psi_q(x)$ and $\Psi(x)$ are the wavefunctions in position representation corresponding to states $|q\rangle$ and $|\Psi\rangle$ respectively.

    A. 1 only  B. 1 and 2 only  C. 1 and 3 only  D. 2 and 3 only  **E. all of the above**

10. Suppose $\{|q_n\rangle, n = 1,2,3 \ldots \infty\}$ forms a complete set of orthonormal eigenstates of an operator $\hat{Q}$ corresponding to a physical observable with non-degenerate eigenvalues $q_n$. $\hat{I}$ is the identity operator. Choose all of the following statements that are correct.

(1) $\sum_n |q_n\rangle\langle q_n| = \hat{I}$ (2) $\langle\Psi|\hat{Q}|\Psi\rangle = \sum_n q_n |\langle q_n|\Psi\rangle|^2$ (3) $\langle\Psi|\hat{Q}|\Psi\rangle = \sum_n q_n \langle q_n|\Psi\rangle$

A. 1 only  B. 2 only  C. 3 only  **D. 1 and 2 only**  E. 1 and 3 only

11. Choose all of the following statements that are correct about the position space and momentum space wavefunctions for this quantum state.

(1) The wavefunction in position representation is $\Psi(x) = \langle x|\Psi\rangle$ where $x$ is a continuous index.

(2) The wavefunction in momentum representation is $\tilde{\Psi}(p) = \langle p|\Psi\rangle$ where $p$ is a continuous index.

(3) The wavefunction in momentum representation is $\tilde{\Psi}(p) = \int dx(-i\hbar\frac{\partial}{\partial x}\Psi(x))$

A. all of the above  B. 2 only  **C. 1 and 2 only**  D. 3 only  E. 1 and 3 only

12. Choose all of the following equations involving the inner product that are correct.

(1) $\langle x|\Psi\rangle = \int x\Psi(x)dx$

(2) $\langle x|\Psi\rangle = \int \delta(x - x')\Psi(x')dx'$

(3) $\langle p|\Psi\rangle = \int \langle p|x\rangle\langle x|\Psi\rangle dx = \int e^{-ipx/\hbar}\Psi(x)dx$ (Ignore normalization issues.)

A. 1 and 2 only  B. 1 and 3 only  **C. 2 and 3 only**  D. 1 only  E. 2 only

13. Choose all of the following statements that are correct.

(1) $|\Psi\rangle = \int \langle p|\Psi\rangle |p\rangle dp$

(2) $|\Psi\rangle = \int \Psi(x)|x\rangle dx$

(3) If you measure the position of the particle in the state $|\Psi\rangle$, the probability of finding the particle between $x$ and $x + dx$ is $|\langle x|\Psi\rangle|^2 dx$.

A. 1 only  B. 1 and 2 only  C. 1 and 3 only  D. 2 and 3 only  **E. all of the above**

14. $|p'\rangle$ is the momentum eigenstate with eigenvalue $p'$ for a particle confined in one spatial dimension. Choose all of the following statements that are correct. Ignore normalization issues.

(1) $\langle p|\hat{p}|p'\rangle = p'\langle p|p'\rangle = p'\delta(p-p')$

(2) $\langle x|\hat{p}|p'\rangle = p'\langle x|p'\rangle = p'e^{ip'x/\hbar}$

(3) $\langle x|\hat{p}|p'\rangle = -i\hbar\frac{\partial}{\partial x}\langle x|p'\rangle = -i\hbar\frac{\partial}{\partial x}e^{ip'x/\hbar}$

**A. all of the above**   B. 1 only   C. 1 and 2 only   D. 1 and 3 only   E. 2 and 3 only

15. Choose all of the following statements that are correct:

    (1) The stationary states refer to the eigenstates of any operator corresponding to any physical observable.

    (2) In an isolated system, if a particle is in a position eigenstate (has a definite value of position) at time $t = 0$, the position of the particle is well-defined at all times $t > 0$.

    (3) In an isolated system, if a system is in an energy eigenstate (it has a definite energy) at time $t = 0$, the energy of the particle is well-defined at all times $t > 0$.

A.   1 only   **B. 3 only**   C. 1 and 3 only   D. 2 and 3 only   E. All of the above

16. Choose all of the following statements that are correct about the time dependence of the expectation value of an observable $Q$ in a state $|\Psi\rangle$. Neither the Hamiltonian $\hat{H}$ nor the operator $\hat{Q}$ depends explicitly on time. (Notation: $\frac{\partial}{\partial t}|\Psi\rangle = \left|\frac{\partial \Psi}{\partial t}\right\rangle$)

    (1) $\frac{d}{dt}\langle Q\rangle = \frac{i}{\hbar}\langle[\hat{H},\hat{Q}]\rangle$

    (2) $\frac{d}{dt}\langle Q\rangle = 0$ in a stationary state for all observables $Q$.

    (3) $\frac{d}{dt}\langle Q\rangle = \left\langle\frac{\partial \Psi}{\partial t}\Big|\hat{Q}\Big|\Psi\right\rangle + \left\langle\Psi\Big|\hat{Q}\Big|\frac{\partial \Psi}{\partial t}\right\rangle$

A. 1 only   B. 2 only   C. 1 and 2 only   D. 1 and 3 only   **E. all of the above**

17. Choose all of the following statements that are <u>necessarily</u> correct.

    (1) An observable whose corresponding time-independent operator commutes with the time-independent Hamiltonian of the system, $\hat{H}$, corresponds to a conserved quantity (constant of motion).

    (2) If an observable $Q$ does not depend explicitly on time, $Q$ is a conserved quantity.

(3) If a quantum system is in an eigenstate of the momentum operator at initial time $t = 0$, momentum is a conserved quantity.

A. 1 only    B. 2 only    C. 3 only    D. 1 and 3 only    E. all of the above

18. Suppose $\{|q_n\rangle, n = 1,2,3 \ldots N\}$ form a complete set of orthonormal eigenstates of an operator $\hat{Q}$ with eigenvalues $q_n$. Which one of the following relations is correct? All of the summations are over all possible positive integer values of $n$ and $m$.

(1) $\hat{Q} = \sum_{n,m} q_n |q_n\rangle\langle q_m|$

(2) $\hat{Q} = \sum_n q_n \langle q_n | q_n \rangle$

(3) $\hat{Q} = \sum_n q_n$

A. 1 only   B. 2 only   C. 3 only   D. 2 and 3 only    **E. none of the above**

19. Hermitian operators $\hat{A}$ and $\hat{B}$ are compatible when the commutator $[\hat{A}, \hat{B}] = 0$ and incompatible when $[\hat{A}, \hat{B}] \neq 0$. Choose all of the following statements that are correct.

(1) You can always find a complete set of simultaneous eigenstates for compatible operators.

(2) You can never find a complete set of simultaneous eigenstates for incompatible operators.

(3) For two compatible operators $\hat{A}$ and $\hat{B}$ whose eigenvalue spectra have no degeneracy, you can infer the value of the observable $B$ after the measurement of the observable $A$ returns a particular value for $A$.

A. 1 only   B. 1 and 2 only   C. 1 and 3 only   D. 2 and 3 only   **E. all of the above**

**For questions 20-30:**

$|s, m_s\rangle$ denotes a simultaneous eigenstate of $\hat{S}^2$ and $\hat{S}_z$ such that the quantum numbers corresponding to $\hat{S}^2$ and $\hat{S}_z$ are $s$ and $m_s$, respectively.

**Raising and lowering operators for spin angular momentum are defined by** $\hat{S}_\pm = \hat{S}_x \pm i\hat{S}_y$ **and** $S_\pm |s, m_s\rangle = \hbar\sqrt{s(s+1) - m_s(m_s \pm 1)}|s, m_s \pm 1\rangle$. **The commutator is defined by** $[S_x, S_y] = i\hbar S_z$.

**For example, for a spin-1/2 particle:**

$$s = \frac{1}{2}, \quad m_s = \pm\frac{1}{2}$$

$$\hat{S}_z \left|\frac{1}{2}, \frac{1}{2}\right\rangle = +\frac{\hbar}{2}\left|\frac{1}{2}, \frac{1}{2}\right\rangle \qquad \hat{S}_z \left|\frac{1}{2}, -\frac{1}{2}\right\rangle = -\frac{\hbar}{2}\left|\frac{1}{2}, -\frac{1}{2}\right\rangle$$

$$\hat{S}_+ \left|\frac{1}{2}, \frac{1}{2}\right\rangle = 0 \qquad \hat{S}_+ \left|\frac{1}{2}, -\frac{1}{2}\right\rangle = \hbar\left|\frac{1}{2}, \frac{1}{2}\right\rangle$$

$$\hat{S}_- \left|\frac{1}{2}, \frac{1}{2}\right\rangle = \hbar\left|\frac{1}{2}, -\frac{1}{2}\right\rangle \qquad \hat{S}_- \left|\frac{1}{2}, -\frac{1}{2}\right\rangle = 0$$

**Eigenstates of** $\hat{S}_x$: $|\uparrow\rangle_x = \dfrac{\left|\frac{1}{2},\frac{1}{2}\right\rangle + \left|\frac{1}{2},-\frac{1}{2}\right\rangle}{\sqrt{2}} \qquad |\downarrow\rangle_x = \dfrac{\left|\frac{1}{2},\frac{1}{2}\right\rangle - \left|\frac{1}{2},-\frac{1}{2}\right\rangle}{\sqrt{2}}$

**Eigenstates of $\hat{S}_y$:** $|\uparrow\rangle_y = \dfrac{\left|\frac{1}{2},\frac{1}{2}\right\rangle + i\left|\frac{1}{2},-\frac{1}{2}\right\rangle}{\sqrt{2}}$   $|\downarrow\rangle_y = \dfrac{\left|\frac{1}{2},\frac{1}{2}\right\rangle - i\left|\frac{1}{2},-\frac{1}{2}\right\rangle}{\sqrt{2}}$

20. For a spin-1/2 particle, suppose $|s, m_s\rangle = \left|\frac{1}{2}, -\frac{1}{2}\right\rangle$ is a simultaneous eigenstate of $\hat{S}^2$ and $\hat{S}_z$ with quantum numbers $s = \frac{1}{2}$, and $m_s = -\frac{1}{2}$. Choose all of the following statements that are correct.

(1) $\hat{S}_+ \left|\frac{1}{2}, -\frac{1}{2}\right\rangle$ is an eigenstate of both $\hat{S}^2$ and $\hat{S}_z$.

(2) If $\hat{S}^2 \left|\frac{1}{2}, -\frac{1}{2}\right\rangle = \frac{3}{4}\hbar^2 \left|\frac{1}{2}, -\frac{1}{2}\right\rangle$, then $\hat{S}_+ \left|\frac{1}{2}, -\frac{1}{2}\right\rangle$ is an eigenstate of $\hat{S}^2$ with eigenvalue $\frac{3}{4}\hbar^2$.

(3) If $\hat{S}_z \left|\frac{1}{2}, -\frac{1}{2}\right\rangle = -\frac{\hbar}{2}\left|\frac{1}{2}, -\frac{1}{2}\right\rangle$, then $\hat{S}_+ \left|\frac{1}{2}, -\frac{1}{2}\right\rangle$ is an eigenstate of $\hat{S}_z$ with eigenvalue $-\frac{\hbar}{2}$.

A. 1 only   **B. 1 and 2 only**   C. 1 and 3 only   D. 2 and 3 only   E. all of the above

21. At time $t = 0$, the initial state of a spin-1/2 particle is $\left|\frac{1}{2}, \frac{1}{2}\right\rangle$ so that $\hat{S}_z \left|\frac{1}{2}, \frac{1}{2}\right\rangle = \frac{\hbar}{2}\left|\frac{1}{2}, \frac{1}{2}\right\rangle$. Choose all of the following statements that are correct at time $t = 0$.

(1) If you measure $S_y$, you will obtain zero with 100% probability.

(2) If you measure $S_z$, you will obtain $\frac{\hbar}{2}$ with 100% probability.

(3) If you measure $S^2$, you will obtain $\frac{3\hbar^2}{4}$ with 100% probability.

A. 1 only   B. 2 only   C. 1 and 3 only   **D. 2 and 3 only**   E. all of the above

22. At time $t = 0$, the initial normalized state of a spin-1/2 particle is $|\chi\rangle = a\left|\frac{1}{2}, \frac{1}{2}\right\rangle + b\left|\frac{1}{2}, -\frac{1}{2}\right\rangle$ where $a$ and $b$ are suitable constants. What is the expectation value $\langle S_x \rangle$ at time $t = 0$?

A. $\langle S_x \rangle = 0$

B. $\langle S_x \rangle = \frac{\hbar}{2}(|a|^2 + |b|^2)$

C. $\langle S_x \rangle = \frac{\hbar}{2}(|a|^2 - |b|^2)$

D. $\langle S_x \rangle = \frac{\hbar}{2}\left(\frac{|a+b|^2}{2} + \frac{|a-b|^2}{2}\right)$

E. $\langle S_x \rangle = \frac{\hbar}{2}\left(\frac{|a+b|^2}{2} - \frac{|a-b|^2}{2}\right)$

**Questions 23-25 refer to Stern-Gerlach experiments with magnetic field gradients in various directions. The neutral silver atoms have zero orbital angular momentum. If an atom in state $\left|\frac{1}{2}, \frac{1}{2}\right\rangle$ (or $\left|\frac{1}{2}, -\frac{1}{2}\right\rangle$) passes through a Stern-Gerlach apparatus with the magnetic field gradient in the negative-z direction (SGZ-), it will be deflected in the +z (or –z) direction, respectively. Assume that the initial state of each silver atom entering the Stern Gerlach apparatus is spatially localized.**

23. A beam of neutral silver atoms, each of which is in a spin state $|\chi\rangle = \frac{1}{\sqrt{2}}\left(\left|\frac{1}{2}, \frac{1}{2}\right\rangle + \left|\frac{1}{2}, -\frac{1}{2}\right\rangle\right)$, propagates into the paper (x-direction). The beam is sent through a Stern Gerlach apparatus with a magnetic field gradient in the −z-direction **(SGZ-)**. Which one of the following schematically represents the pattern you expect to observe on a distant screen in the y-z plane when the atoms hit a screen?

A. 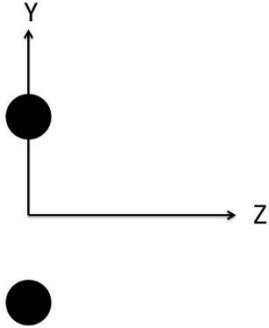

B. 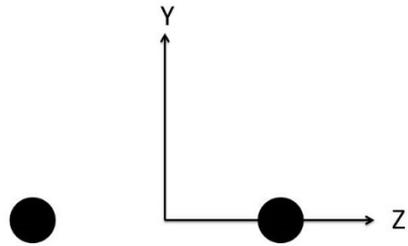

C. 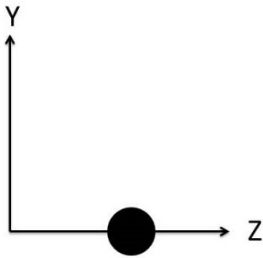

D. 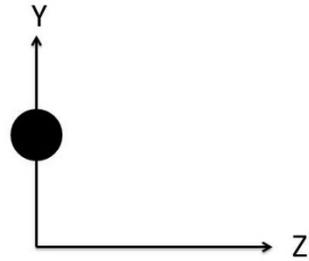

E. None of the above.

24. A beam of neutral silver atoms, each of which is in a spin state state $|\chi\rangle = \left|\frac{1}{2}, \frac{1}{2}\right\rangle$, propagates into the paper (x-direction). The beam is sent through a Stern Gerlach apparatus with a magnetic field gradient in the $-y$-direction (**SGY-**). Which one of the following schematically represents the pattern you expect to observe on a distant screen in the y-z plane when the atoms hit a screen?

A. 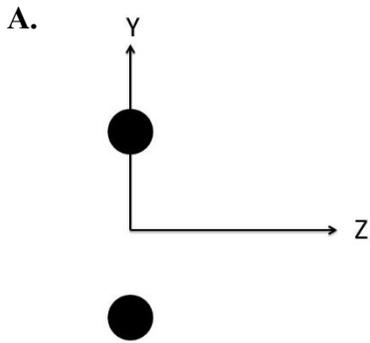
B. 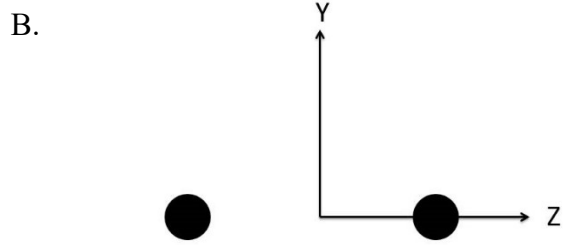
C. 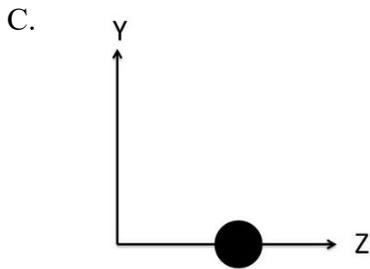
D. 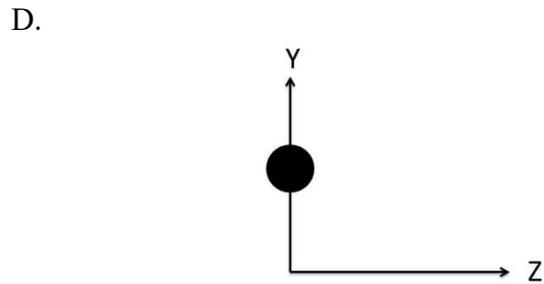

E. None of the above.

25. Suppose Beam A consists of neutral silver atoms, each of which is in the state $|\chi\rangle = \frac{1}{\sqrt{2}}\left(\left|\frac{1}{2}, \frac{1}{2}\right\rangle + \left|\frac{1}{2}, -\frac{1}{2}\right\rangle\right)$, and Beam B consists of a mixture in which half of the silver atoms are each in state $\left|\frac{1}{2}, \frac{1}{2}\right\rangle$ and the other half are each in state $\left|\frac{1}{2}, -\frac{1}{2}\right\rangle$. Both beams propagate along the $y$ direction. Choose all of the following statements that are correct (Note the magnetic field gradient in each case):

(1) The state of each silver atom in beam A will become a superposition of two spatially separated

components after passing through a Stern Gerlach apparatus with a magnetic field gradient in the $-z$-direction **(SGZ-)**.

(2) We can distinguish between Beam A and Beam B by analyzing the pattern on a distant screen after each beam is sent through a Stern Gerlach apparatus with a magnetic field gradient in the $-z$-direction **(SGZ-)**.

(3) We can distinguish between Beam A and Beam B by analyzing the pattern on a distant screen after each beam is sent through a Stern Gerlach apparatus with a magnetic field gradient in the $-x$-direction **(SGX-)**.

A. 1 only    B. 2 only    C. 1 and 2 only    **D. 1 and 3 only**    E. All of the above.

**In questions 26-30, the Hamiltonian of a charged particle with spin-1/2 at rest in an external uniform magnetic field is $\hat{H} = -\gamma B_0 \hat{S}_z$ where the uniform field $B_0$ is along the z-direction and $\gamma$ is the gyromagnetic ratio (a constant). The phrase "immediate succession" implies that the time evolution can be ignored between the first and second measurements.**

26. Suppose that at time $t = 0$, the particle is in an initial normalized spin state $|\chi\rangle = a \left|\frac{1}{2}, \frac{1}{2}\right\rangle + b \left|\frac{1}{2}, -\frac{1}{2}\right\rangle$ where $a$ and $b$ are suitable constants. What is the state of the system after time $t$?

A. $|\chi(t)\rangle = e^{\frac{i\gamma B_0 t}{2}} \left( a \left|\frac{1}{2}, \frac{1}{2}\right\rangle + b \left|\frac{1}{2}, -\frac{1}{2}\right\rangle \right)$

B. $|\chi(t)\rangle = e^{\frac{-i\gamma B_0 t}{2}} \left( a \left|\frac{1}{2}, \frac{1}{2}\right\rangle + b \left|\frac{1}{2}, -\frac{1}{2}\right\rangle \right)$

C. $|\chi(t)\rangle = e^{\frac{i\gamma B_0 t}{2}} \left( (a+b) \left|\frac{1}{2}, \frac{1}{2}\right\rangle + (a-b) \left|\frac{1}{2}, -\frac{1}{2}\right\rangle \right)$

D. $|\chi(t)\rangle = a e^{\frac{i\gamma B_0 t}{2}} \left|\frac{1}{2}, \frac{1}{2}\right\rangle + b e^{\frac{-i\gamma B_0 t}{2}} \left|\frac{1}{2}, -\frac{1}{2}\right\rangle$

E. None of the above.

27. Suppose that at time $t = 0$, the particle is in an initial state in which the $x$-component of spin $S_x$ has a definite value $\frac{\hbar}{2}$. Choose all of the following statements that are correct about measurements performed on the system starting with this initial state at $t = 0$.

(1) If you measure $S_x$ immediately following another measurement of $S_x$ at $t = 0$, both measurements of $S_x$ will yield the same value $\frac{\hbar}{2}$.

(2) If you first measure $\vec{S}^2$ at $t = 0$ and then measure $S_x$ in immediate succession, the measurement

of $S_x$ will yield the value $\frac{\hbar}{2}$ with 100% probability.

(3) If you first measure $S_z$ at $t = 0$ and then measure $S_x$ in immediate succession, the measurement of $S_x$ will yield the value $\frac{\hbar}{2}$ with 100% probability.

A. 1 only  B. 3 only  **C. 1 and 2 only**  D. 1 and 3 only  E. 2 and 3 only

28. Suppose that at time $t = 0$, the particle is in an initial state in which the $x$-component of spin $S_x$ has a definite value $\frac{\hbar}{2}$ (as in the preceding question). Choose all of the following statements that are correct about measurements performed on the system <u>after a long time $t$</u>. (The only difference between the problem statement of questions 29 and 30 is that the measurements are performed at time $t = 0$ in question 29 and after a long time $t$ in question 30.)

(1) If you measure $S_x$ immediately following another measurement of $S_x$, both measurements of $S_x$ will yield the same value $\frac{\hbar}{2}$ with 100% probability.

(2) If you first measure $\vec{S}^2$ and then measure $S_x$ in immediate succession, the measurement of $S_x$ will yield the value $\frac{\hbar}{2}$ with 100% probability.

(3) If you first measure $S_z$ and then measure $S_x$ in immediate succession, the measurement of $S_x$ will yield the value $+\frac{\hbar}{2}$ or $-\frac{\hbar}{2}$ with equal probability

A. 1 only  **B. 3 only**  C. 1 and 2 only  D. 1 and 3 only  E. 2 and 3 only

29. Suppose the particle is initially in an eigenstate of the $x$-component of spin angular momentum operator $\hat{S}_x$. Choose all of the following statements that are correct:

(1) The expectation value $\langle S_x \rangle$ depends on time.
(2) The expectation value $\langle S_y \rangle$ depends on time.
(3) The expectation value $\langle S_z \rangle$ depends on time.

A. 1 only  B. 3 only  **C. 1 and 2 only**  D. 2 and 3 only  E. all of the above

30. Suppose the particle is initially in an eigenstate of the z-component of spin angular momentum $\hat{S}_z$. Choose all of the following statements that are correct:

    (1) The expectation value $\langle S_x \rangle$ depends on time.

    (2) The expectation value $\langle S_y \rangle$ depends on time.

    (3) The expectation value $\langle S_z \rangle$ depends on time.

    **A. none of the above**   B. 1 only   C. 3 only   D. 1 and 2 only   E. all of the above

**A particle interacts with a one-dimensional infinite square well of width $a$ ($V(x) = 0$ for $0 \leq x \leq a$ and $V(x) = +\infty$ otherwise). The stationary state wavefunctions are $\psi_n(x) = \sqrt{\frac{2}{a}} \sin\left(\frac{n\pi x}{a}\right)$ and the allowed energies are $E_n = \frac{n^2 \pi^2 \hbar^2}{2ma^2}$ where $n = 1, 2, 3 \ldots \infty$. Answer questions 31-34.**

31. The wavefunction at time $t = 0$ is $\Psi(x, 0) = Ax(a - x)$ for $0 \leq x \leq a$, where $A$ is a suitable normalization constant. Choose all of the following statements that are correct at time $t = 0$:

    (1) If you measure the position of the particle at time $t = 0$, the probability density for measuring $x$ is $|Ax(a - x)|^2$.

    (2) If you measure the energy of the system at time $t = 0$, the probability of obtaining $E_1$ is $\left| \int_0^a \psi_1^*(x) Ax(a - x)\, dx \right|^2$.

    (3) If you measure the position of the particle at time $t = 0$, the probability of obtaining a value between $x$ and $x + dx$ is $\int_x^{x+dx} x|\Psi(x, 0)|^2\, dx$.

    A. 1 only   B. 3 only   **C. 1 and 2 only**   D. 1 and 3 only   E. All of the above

32. The wavefunction at time $t = 0$ $\Psi(x, 0) = Ax(a - x)$ for $0 \leq x \leq a$, where $A$ is a suitable normalization constant. Choose all of the following statements that are correct at a time $t > 0$:

    (1) If you measure the position of the particle after a time $t$, the probability density for measuring $x$ is $|Ax(a - x)|^2$.

    (2) If you measure the energy of the system after a time $t$, the probability of obtaining $E_1$ is $\left| \int_0^a \psi_1^*(x) Ax(a - x) dx \right|^2$.

    (3) If you measure the position of the particle after a time $t$, the probability of obtaining a value

between $x$ and $x + dx$ is $\int_x^{x+dx} x|\Psi(x,0)|^2 \, dx$.

A. None of the above    B. 1 only    **C. 2 only**    D. 3 only    E. 1 and 3 only

33. The wavefunction at time $t = 0$ is $\Psi(x, 0) = \frac{\psi_1(x)+\psi_2(x)}{\sqrt{2}}$. Choose all of the following statements that are correct <u>at time $t = 0$</u>:

   (1) If you measure the position of the particle at time $t = 0$, the probability density for measuring $x$ is $\left|\frac{\psi_1(x)+\psi_2(x)}{\sqrt{2}}\right|^2$.

   (2) If you measure the energy of the system at time $t = 0$, the probability of obtaining $E_1$ is $\left|\int_0^a \psi_1^*(x) \left(\frac{\psi_1(x)+\psi_2(x)}{\sqrt{2}}\right) dx\right|^2$.

   (3) If you measure the position of the particle at time $t = 0$, the probability of obtaining a value between $x$ and $x + dx$ is $\int_x^{x+dx} x|\Psi(x,0)|^2 \, dx$.

A. 1 only    B. 3 only    **C. 1 and 2 only**    D. 1 and 3 only    E. All of the above

34. The wavefunction at time $t = 0$ is $\Psi(x, 0) = \frac{\psi_1(x)+\psi_2(x)}{\sqrt{2}}$. Choose all of the following statements that are correct <u>at a time $t > 0$</u>:

   (1) If you measure the position of the particle <u>after a time $t$</u>, the probability density for measuring $x$ is $\left|\frac{\psi_1(x)+\psi_2(x)}{\sqrt{2}}\right|^2$.

   (2) If you measure the energy of the system <u>after a time $t$</u>, the probability of obtaining $E_1$ is $\left|\int_0^a \psi_1^*(x) \left(\frac{\psi_1(x)+\psi_2(x)}{\sqrt{2}}\right) dx\right|^2$.

   (3) If you measure the position of the particle <u>after a time $t$</u>, the probability of obtaining a value between $x$ and $x + dx$ is $\int_x^{x+dx} x|\Psi(x,0)|^2 \, dx$.

A. None of the above    B. 1 only    **C. 2 only**    D. 3 only    E. 1 and 3 only